# Synergies between ground-based and space-based observations in the solar system and beyond

*Vincent Kofman [1,2]*
NASA Goddard Space Flight Center
+1 202 957 8685
vincent.kofman@nasa.gov

**Co-authors:**
*Chris Mockel [3], Glenn Orton [4,5], Flaviane Venditti [6,7], Alessandra Migliorini [8], Sara Faggi [1,2],*
*Martin Cordiner [1,9], Giuliano Liuzzi [1,2], Manuela Lippi [1,2], Elise W. Knutsen [1,2,] Imke de Pater [3],*
*Edgard G. Rivera-Valentín [10,11], Dennis Bodewits [12]*
*Stefanie N. Milam [1], Eric Villard [13], Geronimo L. Villanueva [1]*

[1] NASA Goddard Space Flight Center, USA
[2] American University, USA.
[3] University of California, Berkeley, USA
[4] Jet Propulsion Laboratory, USA
[5] California Institute of Technology, USA
[6] NASA Solar System Exploration Research Virtual Institute (SSERVI), USA
[7] Planetary Radar Department, Arecibo Observatory, USA
[8] INAF-IAPS, Istituto di Astrofisica e Planetologia Spaziali, Italy
[9] Catholic University of America, USA,
[10] Lunar and Planetary Institute (LPI), USA
[11] Universities Space Research Association, USA
[12] Auburn University, USA
[13] Joint ALMA Observatory, Chile

**Co-signatories:**
*Nathan X. Roth (GSFC/USRA),*
*Marshall J. Styczinski (University of Washington)*
*Richard J. Cartwright (SETI Institute)*
*Indhu Varatharajan (DLR Institute of Planetary Research)*
*Sarah E. Moran (Johns Hopkins University)*
*Tim I. Michaels (SETI Institute)*
*Alice Lucchetti (INAF-Astronomical Observatory of Padova, Italy)*
*Maurizio Pajola (INAF-Astronomical Observatory of Padova, Italy)*
*Franck Marchis (SETI Institute)*
*Stephen R. Kane (University of California, Riverside)*
*Simone Ieva (INAF - Osservatorio Astronomico di Roma, Italy)*
*Padma Yanamandra-Fisher (Space Science Institute, USA)*

**Introduction** Telescope and detector developments continuously enable deeper and more detailed studies of astronomical objects. Larger collecting areas, improvement in dispersion and detector techniques, and higher sensitivities allow detection of more molecules in a single observation, at lower abundances, resulting in better constraints of the target's physical/chemical conditions. Improvements on current telescopes, and not to mention future observatories, both in space and on the ground, will continue this trend, ever improving our understanding of the Universe. Planetary exploration missions carry instrumentation to unexplored areas, and reveal details impossible to observe from the Earth by performing *in-situ* measurements. Space based observatories allow observations of object at wavelength ranges absorbed by the Earth's atmosphere (i.e. UV, X-ray). The depth of understanding from all of these studies can be greatly enhanced by combining observations: ground-based and space-based, low-resolution and high-resolution, local and global-scale, similar observations over a broader or different spectra range, or by providing temporal information through follow-ups. Combined observations provide context and a broader scope of the studied object, and **in this white paper, we outline a number of studies where observations are synergistically applied to increase the scientific value of both datasets.** Examples include (atmospheric) studies of Venus, Mars, Titan, comets, Jupiter, as well as more specific cases describing synergistic studies in the Juno mission, and ground-based radar studies for near-Earth objects. The examples aim to serve as inspiration for future synergistic observations, and recommendations are made based on the lessons learned from these examples.

**Venus** Venus has been widely investigated from space, thanks to the Magellan and Venera missions in the 70s, the ESA Venus Express and the JAXA mission Akatsuki, the latter in equatorial orbit around the planet since December 2015. Ground-based radar studies of Venus provided our early understanding of its surface geology, such as the paucity of impact craters and the abundance of volcanic landforms. Furthermore, ground-based observations have contributed significantly by being able to monitor the complex dynamics of Venus and study the wind fields at different altitudes. Results are used as a background and reference for the similar measurements acquired with remote sensing instrumentations (*1, 2*). Trace gases, such as $H_2O$, HDO, CO, $SO_2$ can be studied using ground-based facilities, in spectral regions that are not covered with spacecraft instrumentation (*3, 4*). Indeed, the infrared and submillimeter spectral ranges are populated with particularly strong features of trace species present in Venus' atmosphere that can be easily identified, and from the abundances and their evolution and we can infer the dynamics of the atmosphere, explore the driving factors that control the climate, allowing to test our understanding of global circulation and the validity of global circulation models.

**Mars** Due to its proximity, Mars is a prime target for *in-situ* missions, both on the ground and in orbit, and can be studied using Earth ground-based, spatially-resolved observations, logically leveraging all of these approaches to enhance our understanding of the planet. The study of trace gases such as $H_2O$/HDO (*5–7*) $CH_4$, (*8–11*) and $C_2H_6$ (*8, 12*) in the Martian atmosphere serve as excellent examples where measurements from rovers, orbiters and Earth-based studies are synergistically combined to yield a depth of understanding far beyond what the separate measurements could yield. The distribution of trace gases in the atmosphere of Mars are of interest to understand the potential for (past) habitability and to gain insight into geologic processes occurring on or below the surface. Recently, the capability of monitoring trace gases, their isotopes, and atmospheric escape have been greatly enhanced by two dedicated *in-situ* missions, the European ExoMars Trace Gas Orbiter (TGO) and NASA Martian Volatiles and EvolutioN (MAVEN). The instruments onboard the two orbiters have the capability to obtain local, high-precision measurements of the vertical structure of the atmosphere as far as thermal structure, water vapor and its isotopes (*6*), aerosols (*13*) trace gases and atmospheric escape (*14*) are concerned. On the other hand, Earth-based high-resolution spectroscopy allows sampling the full observable Mars disk, enabling the study of short-term phenomena, diurnal processes (across the East-West axis) and exchange/circulation between the hemispheres, or even map (sub-surface) geological features using ground-based radar (*15, 16*). As such, Earth-based observations can provide spatial and temporal context and the framework to interpret and extrapolate the results from the *in-situ* vertical profiles. In addition, the rich observation history of Mars, and the availability of successful global climate models provides a framework for the interpretation of observations.

**Titan** Remote and *in-situ* measurements of Titan's atmosphere have identified a wealth of complex organic molecules, the abundances, distributions and temperatures of which are influenced by seasonally variable insolation and global circulation patterns. Studies of Titan can therefore provide unique insights into fundamental (e.g. seasonal) atmospheric processes such as photochemistry, cloud and haze production and winds in a more strongly reducing environment than found on Earth and the other terrestrial planets (*17*). With the advent of ALMA, sub-mm interferometry has emerged as a powerful ground-based technique to study Titan's molecular distributions (*18, 19*) from the Earth's surface. This has enabled instantaneous global mapping of key nitriles (R-CN containing molecules) and their isotopologues (*20*), as well as the first spectroscopic detections of several new organic molecules (*21, 22*). Cassini's mass spectrometry, on the other hand, gave us unique information on the abundances of a wide range of molecules, but with limited spatial coverage, and insufficient mass resolution for unambiguous detections in some cases. Meanwhile, the Cassini Infrared Spectrometer (CIRS) provided extremely sensitive, high-resolution mapping of vibrational emission from Titan's hydrocarbons. Through a synergistic approach involving ground-based and in-situ techniques, a complete picture of Titan's atmospheric chemistry and dynamics was revealed. The limited field-of-view, high-resolution spatial coverage of Cassini CIRS was combined with ALMA global snapshot imagery to piece together the molecular abundances as a function of altitude, enabling rigorous tests for chemical models (*18, 23*). This has resulted in new constraints on the photochemical pathways towards complex organics in primitive planetary atmospheres, leading to an improved understanding of the interplay between photochemistry and global circulation. The Cassini mission was active during 2004-2017, so ground-based mm/sub-mm spectroscopy will be crucial for continuing Cassini's legacy of detailed atmospheric studies, in order to form a more complete understanding of the chemical and climatological variations that take place over a full (29.7 yr) Titan year.

**Comets** Both space-based and ground-based observations have played critical roles in the understanding of comets (for an outlook, see white paper Roth et al., 2020). Ground based observations provide essential context as it allows to place spacecraft targets in the context of the wider population of small bodies, and can provide long-term, temporal context of the behavior and the evolution of mission targets. Visiting a comet offers unique insights to the comet's nucleus, and enables probing the coma composition with instrumentation sensitive to a plethora of molecules different from those detectable to remote observations. In addition, *in-situ* observations allow localized studies of the activity and the chemical and physical processes occurring on the surface and in the inner coma (*24*). For example, the recent ESA/Rosetta mission resulted in a number of paradigm shifts regarding our understanding of comets, including the presence of noble gasses, the discovery of large amounts of $O_2$, and a strong variation in $CO_2$ versus $H_2O$ dominated sublimation (*25–29*). Similar to the other examples, the synergy between space-based and ground-based observations provides complementary information and significantly improved characterization of comets. Ground-based observations of active comets at different wavelengths play an essential role in quantifying the overall coma activity and the chemical composition of the nucleus. In particular, for the near-IR spectral region observations of comets benefit from (1) spectral high-resolution, needed to discern the ro-vibrational structure of the observed molecules, (2) cross-dispersion, in order to observe different species at the same time, (3) long slits which enables to study the spatial profile and identify outgassing asymmetries in the coma, (4) large collecting area and AO (Adaptive Optics) modules, to improve the sensitivity and contrast and allow the exploration of fragments and jets surrounding the nucleus, (5) a telescope-guiding system, since comets are fast moving objects. In the 3 to 5 μm window, many organic species and their isotopes have been investigated in the past decades (*30–39*).

A powerful example of how the combination from different investigations can provide important discoveries is the discovery of salts as a source for volatiles. Ground-based studies indicated that disrupting comets showed a much higher mixing ratio of HCN and $NH_3$ relative to $C_2H_6$ than other comets, suggesting the existence of a common progenitor such as the ammonium cyanide salts (*40*). This prompted the ROSINA team to search for these species in the coma of 67P, and the subsequent detection of five ammonium salts (*41*), confirming such salts as an important new component of cometary nuclei.

Finally, space-based observations have been essential in quantifying $CO_2$ emission from comets, which is not observable from the ground due to the significant telluric contamination. The Akari Space Observatory measured combined $CO/CO_2$ emissions, but synergistic ground-based high-resolution investigations were required to quantify the CO contribution and accurately quantitatively constrain the $CO_2$ production rates (*42*).

**Jupiter** Figure 1 exemplifies how different observatories are sensitive to various aspects of the atmosphere of Jupiter, with increasing wavelengths allowing to probe deeper into the planet.

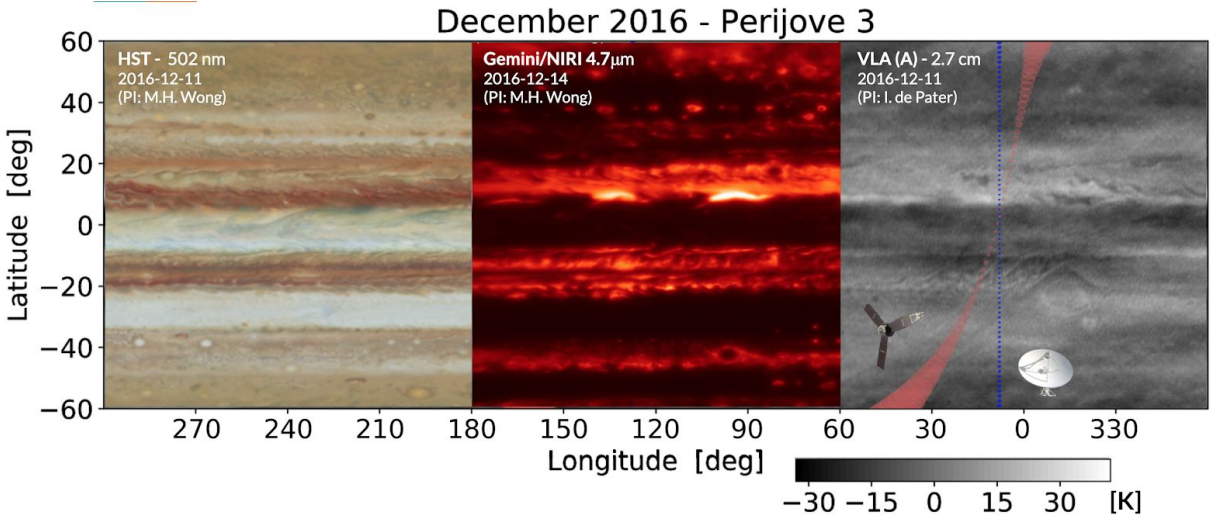

*Figure 1: Composite map of Jupiter, showing visible, infrared and radio images (43). In the right panel, blue and red ovals indicate the spatial scale of VLA and Juno's MWR respectively.*

The image in the left panel (Hubble Space Telescope, visible-light composite, (*44*)) displays the reflected light from the top of the clouds and represents the familiar picture of the zones (white) and belts (brown), where the zones are regions where trace gases are upwelled to create a thick $NH_3$ cloud deck, and the belts are regions with less cloud coverage due to the downwelling of dry air. The HST image shows the complex interplay of winds, jets and storms and constrains the upper boundary conditions for the atmospheric retrievals. In the infrared (Gemini, 4.7 μm), shown in the middle panel, the brighter regions correspond to the darker regions in the visible image, showing thermal emission from deeper, warmer parts of the atmosphere (*44*). In the right panel, residual temperatures of radio observations (Very Large Array, image at 2.7 cm) are shown, where brightness temperature trace variations in the abundance of ammonia gas as a function of depth (*45*). Combining the visible, infrared, and radio images allows for a much better understanding of distribution and motions of different gaseous constituents. The atmospheric dynamics probed across the troposphere gives clues on the processes that shape the climate on Jupiter, such as the interplay of convection and planetary waves that cause planetary-scale storms (*46*). While the ground-based observations map the entire planet down to

~10 bar (blue shaded region corresponds to the VLA resolution), the Juno Microwave Radiometer (MWR) measurements shown as the red shaded regions on the radio map, cover 1.3-50 cm sensing thermal emission much deeper than the VLA – all way down to hundreds of bars (*46*), but cover a much smaller region of the planet. Juno detected lightning strikes, linking the high-energy storms with formation convective cells over deep water clouds (*44, 48*).

**Juno's Earth-Based Observational Program** The Juno mission serves as a paradigm for strong synergy between Earth-based observations and interplanetary data. Within the mission, a large role is played by a coordinated set of Earth-based (both ground-based and Earth-proximal) observations that substantially extend and enhance the scientific results returned by the mission. This is manifested in several distinct ways. First, Juno's instruments are extremely close to the planet, resulting in extraordinary spatial resolution, but missing contextual information (Figure 1, right panel). This leads to the question whether the regions measured are similar to others on the planet, or are they unique? Are they a part of a phenomenon affecting a much broader area, such as planetary-scale waves? Second, Juno's observations sense the atmosphere and aurora in a very limited time, a "snapshot" of a very active planet. Observations throughout the mission provide a context in time to track the evolution of specific features Juno is measuring. Third, Juno is not equipped with instruments that include the full spectrum of radiation emerging from the atmosphere, and Earth-based observations supply key parts of the spectrum that are not measured by Juno's complement of instruments, such as the X-ray and mid-infrared. Finally, Earth-based observations can make simultaneous or contemporaneous observations of multiple components of the Jovian system that affect the interpretation of the results its own instruments provide, *e.g.* the state of active volcanism on the satellite Io and the extent to which it is injecting particles into the magnetosphere. The concerted observations allow a substantial enhancement of mission results. For the case of the Juno mission, over 40 papers involving Juno-supporting observations, often combined with Juno observations, have been published in the peer-reviewed literature by this time.

**Ground-based observations of small bodies** Ground or space-based optical light-curves can provide constraints on the rotation rate and orientation of small bodies, and can support shape reconstruction in the case of well resolved targets. Infrared spectra can be used to determine the albedo and thermal inertia, constrain the surface roughness, and reveal changes in thermal characteristics with rotation. Radar observations can provide information about shape, spin, size, surface properties, near-surface geology, and reveal natural satellites (Virkki et al., 2020). Combined optical and space infrared near-Earth asteroid (NEA) surveys allow for rapid post-discovery radar measurements that significantly decrease the orbital parameters uncertainties and thus extend the accuracy and time frame of ephemeris calculations (*49*). Ground-based radar, thermal spectroscopy, and optical light curves are combined for a better characterization of NEAs, and when combined can be used for a more precise construction of a

3D shape model (*50–52*). Ground-based radar facilities also play a major role in small-body (i.e. asteroids, and comets) astrometry and characterization, planetary defense (e.g., see white papers Mainzer et al., 2020; Virkki et al., 2020 and *53*), but can be applied to or many more objects (e.g. white paper Rivera-Valentín et al., 2020). The most powerful facility for ground based planetary radar is the Arecibo Observatory in Puerto Rico, operating in S-band (2.38 GHz, 1 MW), and is heavily used for follow up of near-Earth objects to support planetary defense and small bodies missions (*i.e.* NASA's DART mission, JAXA's Hayabusa, Destiny+, etc), as well as the Goldstone Solar System Radar in California.

**Findings and recommendations** Joint observations of interplanetary targets by both spacecraft and Earth-based instruments have provided and continue to promise an enriched level of science products emerging from the investigations. A number of examples were described that highlight the value of combining ground-based with space-based observatories and/or local spacecraft measurements. Even observations from ground-based observatories at different wavelength ranges hold the key to reveal deeper understanding when these can be combined.

**Summary of findings:**
• Synergistic observations on Mars between trace gas orbiters, rovers, and high-resolution Earth-based studies enable detailed mapping trace gas species. In particular, vertical profiles and maps are combined to yield deep insight into atmospheric processes taking place on the planet.
• Ground-based observatories and radar facilities play key roles in space mission support both prior and during missions from a 1) navigational (i.e. ephemeris and landing locations), 2) scientific perspective (i.e. larger context of mission observations, concurrent observations of different molecules/physical structures, or guide in-situ measurements).
• Multi-wavelength (i.e. visible, near/mid-infrared and radio) yield different windows into the same object, and when the observations can be combined yield significantly more scientific return as the chemical/physical parameter space can be narrowed.
• The combination of different observational facilities may yield unexpected new discoveries, invigoration research and resulting in exciting new insights into astronomical objects.

**Recommendations:**
• Strong ground-based support programs enrich scientific return of space missions, and do not necessarily require state-of-the-art facilities. Although NASA's Solar System Observations (SSO) currently support proposals aimed at this, dedicated time for coordination from scientists involved in the mission is essential, as these are immensely beneficial to the total science return of the mission. Strategic planning should consider the importance of dedicating funds to support programs.
• Scientist and proposal selection procedures should consider and prioritize simultaneous multi-observatory observations; both ground and space-based observatory planning should strongly consider coordination concurrent observations.

Looking towards the future of planetary science to the exploration of planetary atmospheres beyond the Solar System, multiwavelength observations will be key to characterize the atmospheres of exoplanets through direct imaging and transmission spectroscopy. Solar system observations provide the framework on how to best combine observations and provide the ground truth to increase the fidelity of exoplanet atmospheric retrievals.